# Peripheral brain interfacing: Reading high-frequency brain signals from the output of the nervous system

Jaime Ibáñez[1,2*], Blanka Zicher[3], Etienne Burdet[3], Stuart N. Baker[4], Carsten Mehring[5], Dario Farina[3,*]

[1] BSICoS, I3A, IIS Aragón, Universidad de Zaragoza, Zaragoza, Spain

[2] Centro de Investigación Biomédica en Red en Bioingeniería, Biomateriales y Nanomedicina (CIBER-BBN), Instituto de Salud Carlos III, Madrid, Spain

[3] Department of Bioengineering, Imperial College, London, United Kingdom

[4] Institute of Neuroscience, Newcastle University, Newcastle upon Tyne, UK

[5] Faculty of Biology, University of Freiburg, Freiburg im Breisgau, 79104, Germany

* Corresponding Authors: jibanez@unizar.es, d.farina@imperial.ac.uk

// ABSTRACT //

Accurate and robust recording and decoding from the central nervous system (CNS) is essential for advances in human-machine interfacing. However, technologies used to directly measure CNS activity are limited by their resolution, sensitivity to interferences, and invasiveness. Advances in muscle recordings and deep learning allow us to decode the spiking activity of spinal motor neurons (MNs) in real time and with high accuracy. MNs represent the motor output layer of the CNS, receiving and sampling signals originating in different regions in the nervous system, and generating the neural commands that control muscles. The input signals to MNs can be estimated from the MN outputs. Here we argue that peripheral neural interfaces using muscle sensors represent a promising, non-invasive approach to estimate some neural activity from the CNS that reaches the MNs but does not directly modulate force production. We also discuss the evidence supporting this concept, and the necessary advances to consolidate and test MN-based CNS interfaces in controlled and real-world settings.

// MAIN TEXT //

*Recording neural activity from the brain and spinal cord*

To produce movement, neural signals from the central nervous system (CNS) reach the muscles and control when and how they produce forces. The motor unit is the key structure in the translation of neural information into muscle contraction.

A motor unit consists of a spinal motor neuron (MN) and the group of muscle fibres that it innervates and controls[1]. Motor units represent the frontier between neural processing and behaviour, and MNs are the last common neural pathway controlling behaviour[2]. The activation of each MN in the spinal cord is mirrored by the activation of the innervated muscle unit, thus the muscle unit can serve as an accessible proxy of neural activity. From an engineering perspective, reading the activity of MNs provides access to the output of the somatic motor nervous system (Fig. 1) and a way to infer its working principles.

Currently, methods for directly (i.e., invasively) recording representative ongoing activity from the CNS are still scarce. However, there is growing evidence of increasingly safe approaches, such as intracranial recordings utilized in various medical devices, which have led to significant advancements in fields like speech prostheses, and other types of brain-machine interfaces for patients with severe paralysis[3,4]. Nonetheless, exploring alternative methods for recording CNS activity remains crucial to address existing limitations in current technologies and to further enhance safety and efficacy in neurotechnologies[5]. As a potential alternative to direct brain recordings, we can use an inverse modelling approach to look into some signals in the CNS by decoding the final electrical activity produced at its outputs, in the periphery of the nervous system.

Recent advances in sensors and signal processing methods, including deep learning methods, have made it possible to access the neural activity of sets of MNs by recording the activity they evoke in muscles using electromyography (EMG)[6]. EMG recordings represent the mixture of the electrical activity of the muscle units composing a muscle, which are the proxy of the MNs in the spinal cord. EMG signals are therefore recordings that embed the final output of the CNS. Identifying motor unit activity in the EMG signals then yields a neural interface. This concept is schematically described in Fig. 1A. In this scheme, the CNS and peripheral nervous system constitute a large neural network with high complexity. This network projects its activity to MNs, which form the last layer of the network and have connections with different parts of the CNS (Fig. 1B)[7].

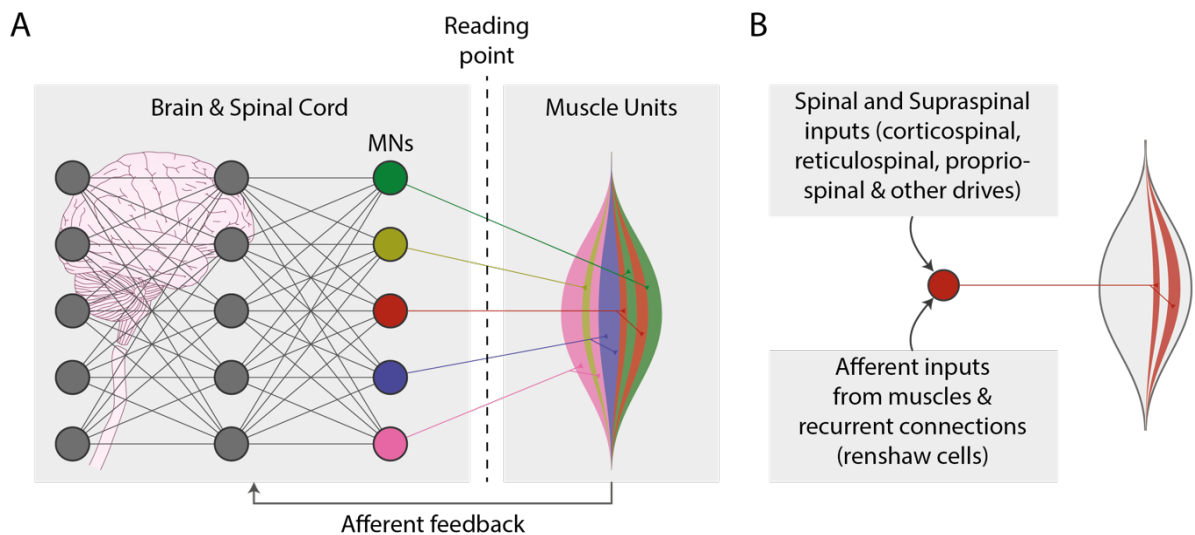

**Fig. 1. Convergent inputs to motor neurons.** (A) MNs in the muscles represent the output layer of the CNS. By measuring the electrical activity that muscles produce during contraction, we can characterize the firing patterns of MNs and therefore "read" the neural activity of this last layer of the nervous system complex network. (B) MNs receive inputs from multiple sites in the CNS (including direct brain projections) and from sensory organs in the periphery. All this information drives MN activity[7].

By characterising MN activity reaching muscles, we access some of the neural signals most tightly associated with behaviour. This has been a key reason for the use of myocontrol for interfaces in clinical applications, such as using EMG to estimate motion intent and to control active prosthetics[8]. However, is the motor command to the muscles (that is, the neural inputs to muscles that determine or modulate the level of contraction) the only information that MNs encode? Or does the MN activity also encode activity in the CNS that does not directly modulate muscle force? This question relates to how much information on the neural processing within the CNS we can extract from MN recordings.

As MNs control muscles, one would think that their output is tightly matched with the muscle characteristics so that the full information transfer from MNs to muscles is associated with force production. However, numerous studies have shown that MN discharge patterns, and therefore EMG signals, contain information that is related to brain oscillations but not directly to force production. For example, it is well known that cortical and EMG signals are correlated in the beta band (13-30 Hz)[9–11], which lies outside the muscle contraction dynamics[12,13] and, thus, information in this band cannot influence force directly, at least within a linear approximation of force production.

These classical observations raise questions about which cortical oscillations are transmitted to the output of the CNS (the MNs) and why. They also suggest that part of the neural activity generated in the brain and containing relevant information regarding brain states and dynamics may be observed in the electric fields produced by muscles, even if it is not associated with an underlying muscle contractile activity. This is because there is a mismatch between the bandwidth of the electrical and mechanical activity of muscles: muscle electric signals have a broader spectral bandwidth than mechanical muscle contraction and thus potentially contain more information. EMG therefore may offer a window through which we can look at brain activity.

This perspective discusses existing evidence supporting the possibility of using the activity of MN populations to estimate neural activity originating in the CNS and not directly involved in force production. This introduces the novel concept of brain interfacing via muscle sensors, which differs from classic EMG-based interfaces aimed at decoding the motor command sent to the muscles to generate an intended action or gesture. In this line, we envision a technology that senses the electric fields in muscles and processes it to infer informative brain signals such as, for example, circuit-specific cortical oscillations at different frequencies in patients with a brain lesion, or changes in identified brain signals that can be used to develop novel paradigms and strategies for human-machine interaction.

The following sections introduce the key concepts that allow the decomposition of large populations of MNs using non-invasive or minimally invasive sensors, and the physiological basis supporting the notion that this neural information can be used to infer ongoing neural activity in relevant regions of the CNS, such as the motor cortex. Finally, we propose frameworks in which MN-derived information can be used to advance the fields of neuroimaging, neural interfacing, and the understanding of neural control of movement.

*Muscle interfaces to decode the output of the CNS*

The contribution of a motor unit to the electric potential recorded at a certain point on a muscle can be modelled as a train of delta functions convolved with the motor unit-specific action potential waveform, as measured from the recording point[14]. The EMG recording thus corresponds to the sum of the electric potentials produced by all the active motor units and noise components.

In this model, the trains of delta functions associated with each motor unit are the neural activity (*i.e.*, the output of the CNS) which can be estimated through the deconvolution of the EMG (Fig. 2). The possibility of identifying this neural activity from the EMG signals depends on two key properties of the biological and biophysical systems underlying muscle activity:

1. The stability of the synaptic connection between each MN and its muscle unit. This connection means that each action potential travelling along the axon of an MN determines a corresponding action potential in the innervated muscle fibres, i.e. in the corresponding muscle unit[15]. This one-to-one relation between MNs and muscle unit action potentials implies that the spiking activity of single neurons can be measured by identifying the action potentials produced by muscle fibres of a motor unit.
2. The spatial and temporal differentiation of the electrical activity that muscle fibres produce in the surrounding volume conductor, i.e. the motor unit electrical signatures. Each motor unit is unique within the muscle in terms of the spatial distribution of muscle fibres, neuromuscular junctions, and fibre membrane characteristics. This implies that the convolutive filters are unique for each motor unit, at least under ideal conditions of sampling in space and time.

Within a linear and stationary model as that shown in Fig. 2, the above two properties and the sparsity of spiking of individual MNs allow us to solve the deconvolution problem and to isolate the activity of individual MNs[6,16].

The deconvolution of neural activity from invasive or non-invasive EMG recordings can be solved by blind source separation methods customised to the convolutive nature of the mixtures and the sparsity of the sources (Fig. 2). Alternatively, deep neural networks can be employed[17] or even generative models of EMG and their inverse formulation[18,19]. The current main aim in the development of more refined methods for EMG deconvolution is the increase in the number of decoded MNs, which can be attained both by refinement in the electrode technology and by advances in the decoding algorithms[20].

The performance of current conventional methods to decode EMG strongly depends on the number of neural sources relative to the number of observations, and on the stability of the action potential waveforms of individual motor units. Because of the large number of motor units in a muscle, only a part of them can be accurately identified by current source separation methods. The decoding is typically limited to the sources with the greatest energy and with distinctive motor unit action potential shapes (if two motor units produce action potentials with comparable temporal-spatial shapes as measured from the recording points, they will not be distinguishable)[21]. In addition, the ability to decompose sources is limited in all contractions that involve changes in joint angles, *i.e.*, most natural tasks. This is because the change in muscle length and relative position with respect to the skin with motion modify the characteristics of the volume conductor separating the sources and the recording electrodes[22,23].

AI-based approaches may be used to mitigate the above limitations. An improvement over conventional statistical source separation methods can be obtained by training deep networks to perform the decomposition process using deconvolution methods for labelling and then increasing the robustness of the network-based decomposition by extensive data augmentation. Interestingly, current models of EMG generation are sufficiently realistic to provide simulated data useful for augmenting neural network-based decomposition of experimental data. For example, recent studies have shown that a realistic model of EMG generation can be used to augment the training of a network based on gate recurrent units (GRU) that learn how to decompose EMG signals so that the resultant decomposition becomes more robust than the original deconvolution algorithm used for labelling[24]. Moreover, it has been recently shown that a hybrid architecture that combines elements of deep latent variable models and conditional adversarial training can provide a generative model of EMG that can mimic the outputs of complex numerical models of human muscle biophysics. Such framework can be used to continuously sample from a dynamically changing system in real-time[18]. This allows us, for the first time, to model all factors of variability in the generation of EMG signals and use them for data augmentation, as described above, as well as to derive inverse models that estimate the internal parameters (such as the MN discharge patterns) based on observations.

Finally, the decomposition process can be performed in a fully unsupervised way by identifying the embedded neural code in a reduced dimensional space, imposing constraints on the properties of this space. For example, Mayer et al. recently proposed an autoencoder that learns the discharge times of individual motor units by processing multichannel EMG recordings[25]. Following pre-processing of the EMG via convolutive sphering, the encoder was constrained to apply an orthogonal transformation to the spatial data and the optimisation criterion enforced a temporal sparsity constraint. Because of the encoding/decoding structure, the method does not need to be trained based on known decomposition results.

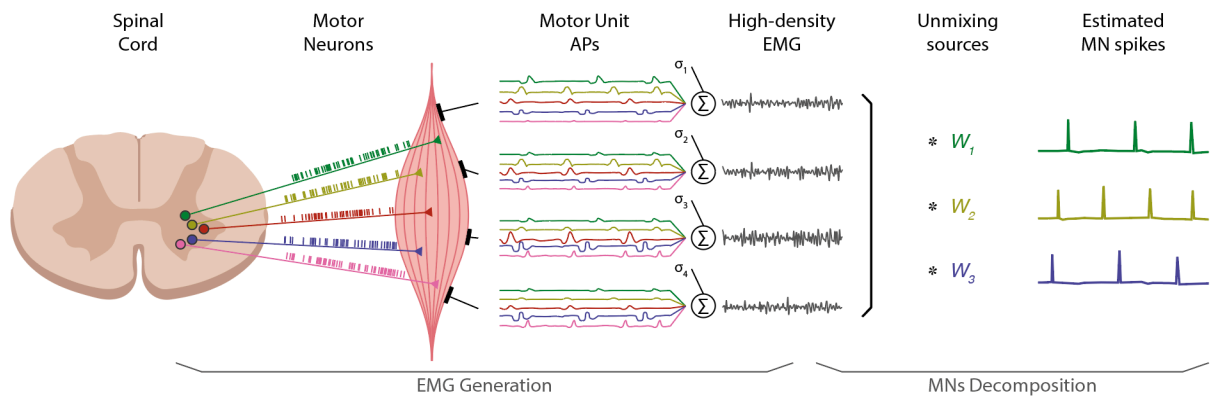

**Fig. 2. Simplified model of EMG generation and the decomposition of MN activity from multichannel EMG with classic deconvolution methods.** MNs have their cell body in the ventral horn of the spinal cord. Based on the synaptic inputs they receive they generate trains of action potentials (APs) that travel along their axons to the muscles. Each MN AP causes a corresponding muscle fibre AP that travels along the muscle fibres of the motor unit formed with the innervating MN. Placing EMG electrodes at different points on the muscle results in recordings of the generated muscle fibre APs by different units with different waveforms (based on the relative position of the electrode and the recorded fibres) plus additional independent noise of standard deviation σ. By multiplying the EMG recordings by unmixing matrices, we can enhance the time instants at which specific motor units are generating action potentials travelling along the muscle fibres. This provides a reliable readout of the spiking activity of MNs corresponding to the identified motor units.

*MNs linearly transmit high-frequency neural signals*

Spinal MNs receive a barrage of synaptic inputs from afferent nerves coding proprioceptive and other sensory information, as well as from spinal and supraspinal circuits, including direct inputs from the cerebral cortex through monosynaptic corticospinal projections[26]. All these inputs are integrated and translated into spiking activity transmitted to muscles (Fig. 1b). Therefore, the twitches that muscle fibres produce can be seen as the result of a complex transformation of the net excitatory inputs that MNs receive from different sources in the CNS. The CNS then has to ensure that the appropriate neural inputs are received by MNs to generate an intended movement.

The net excitatory input received by MNs is typically modelled as a set of independent and common synaptic inputs[27]. Independent inputs represent the synaptic inputs to MNs that are not correlated with the inputs received by other MNs in a pool. Conversely, common synaptic inputs are those that are the same or correlated to inputs that (many) other MNs in the pool receive[28–30]. Due to the differences in the biophysical properties of MNs within a pool, the way in which they react to inputs tends to linearize and amplify the common inputs with respect to the independent inputs when the neural contents in the MNs are analysed at the population level[31–35]. For this reason, a group of MNs receiving and sampling common and independent inputs can be viewed as a neural linear amplifier of the common inputs with a very high rejection of independent noise[36]. This is evident by considering the layer of MNs as a single-layer perceptron that acts as a linear regressor if the activation functions are linear. In the biological case, this corresponds to considering the MNs as active approximately in the middle of their input-output response curve[37]. We can therefore conclude that the last layer of the CNS is an approximately linear layer that amplifies the common inputs it receives when considering the activity of populations of MNs fully recruited and not reaching saturation[38]. The amount of amplification of the linear components by a group of MNs will increase as a function of the group size[36]. Indeed, the observation that common inputs to MNs acts as a linearization process aligns with population coding principles also observed elsewhere in the nervous system[39,40]. For instance, in the retina, decoding the activity of small populations of retinal ganglion cells rapidly saturates information extraction, indicating that a few neurons can represent the encoded information effectively when this is distributed commonly[41]. Similar results have been observed in the auditory system using multiple spike trains in auditory receptor neurons[42].

Importantly, the linearization done by MNs applies to any common inputs received by MNs: groups of MNs can effectively transmit multiple sources of information at different frequencies simultaneously[35]. Fig. 3 exemplifies this property of MNs in a simulation in which a group of MNs receives common inputs at different frequencies and independent inputs (different for each MN). Simple band-pass filtering of the sum of the spike trains of all MNs allows a reliable estimation and separation of the common inputs at multiple frequencies.

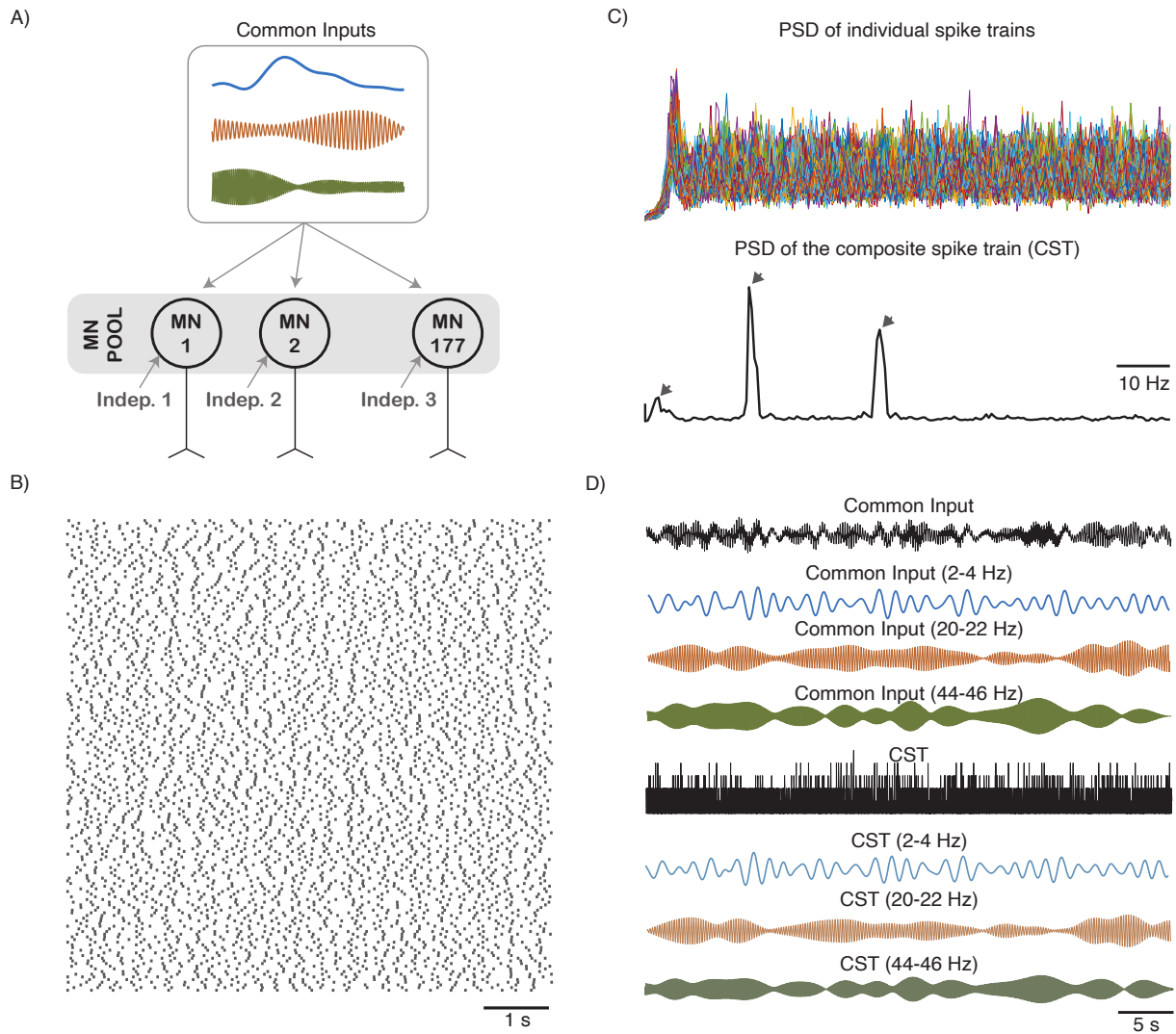

**Fig. 3. Linearization of common input transmission by MNs.** (A) Simulation of a group of 177 motor neurons receiving a common input (sum of three inputs at 2-4 Hz, 21 Hz and 45 Hz) plus an independent input which is different for each MN. (B) Spike trains of a subset of the simulated MNs. (C) Power spectral densities (PSD) of the spike trains produced by the individual MNs (top) and of the composite spike train (sum of all spike trains). The mean of the signals was subtracted to obtain the PSD. (D) The common input to the MNs and the band-pass filtered versions of the common input at different frequencies (top four plots) and the resulting cumulative spike train (CST) and its filtered version at the frequencies of the components contributing to the common inputs. To simulate the results in this figure, a previously validated model was used[27,43]. The model simulated the spiking activity of 177 MNs driven by common and independent inputs. The common input was a signal obtained by summing white gaussian noise filtered in three bands (2-4 Hz, 20-22 Hz and 44-46 Hz). The three filtered signals were normalised in power before summing them. The common input also included a DC component so that the simulated number of inputs arriving from the common projection per 0.2 ms time step was 1.7 (which is equivalent to each MN receiving 8500 impulses/s). In addition to the common input, MNs also received independent inputs which were modelled as white gaussian noise with mean equal to variance. The level of the independent input was adjusted to make MNs fire at physiologically realistic levels during low force contractions.

The conversion from motor unit action potentials into force by muscles can be modelled as a second order critically damped filter, strongly attenuating high-frequency neural inputs to muscles (here we consider high-frequency inputs to muscles those above ~10 Hz)[12,13,44,45]. Indeed, the frequency of movements and forces produced by limbs are constrained to rather low frequencies, with (pathological) tremors representing the highest frequencies that can be measured in the arms and hands [46,47]. However, and somewhat surprisingly, the neural inputs that MNs receive, and therefore their outputs, go far beyond the maximal working frequencies of muscles. This can be experimentally observed when decoding a large number of MNs that can show common activity at frequencies >70-80 Hz[20]. This neural information may consist of different components covering different frequency ranges and originating in different regions of the nervous system which project onto the MN pools. Given the characteristics of muscle transformation of neural inputs into forces, this wide spectrum of common inputs suggests that a significant portion of common synaptic inputs to MNs will be transmitted along their axons without having a direct influence on muscle forces (Fig. 4). These high-frequency inputs may be associated with ongoing activity in the CNS reaching and modulating the firing patterns of MNs. Because of the linear transmission by MN populations, the extraction of this high-frequency activity, which is neural activity that does not affect forces directly, can be achieved using standard linear separation methods, as shown in Fig. 3.

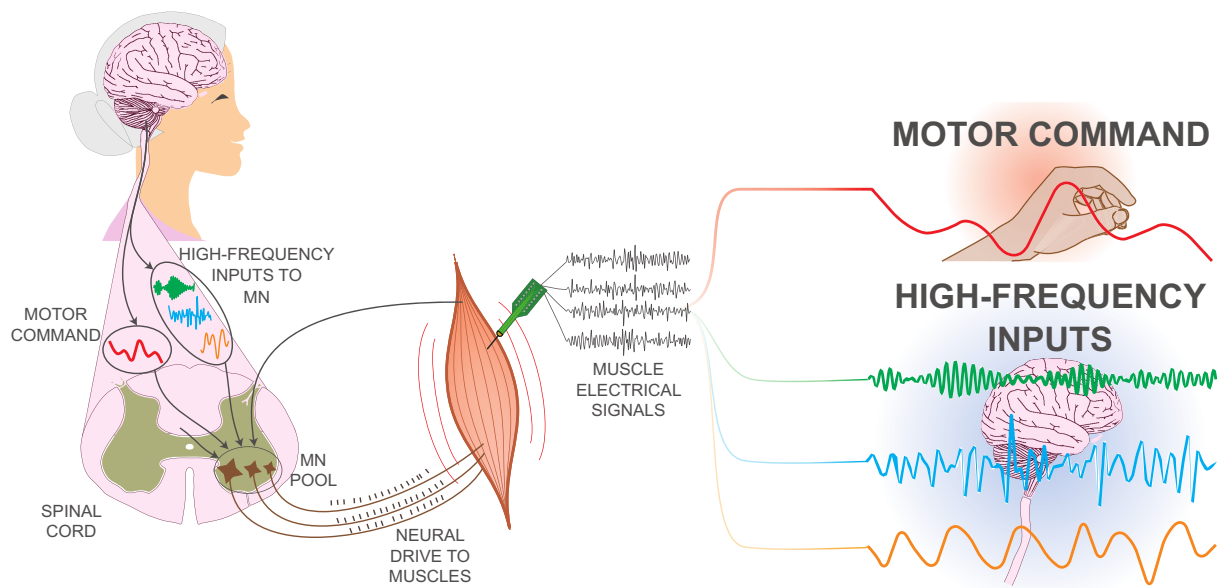

**Fig. 4. Separation of common synaptic inputs to MNs.** MN pools can sample multiple common synaptic inputs. A portion of these inputs (the low-frequency inputs) control the level of muscle contraction to determine movement [12,44,45]. The rest of the inputs (high-frequency inputs) do not interfere with movement directly but are still encoded by the trains of action potentials travelling along MNs and muscle fibres and can be decoded.

*Accessing brain activity through MNs*

Identifying the sources contributing to the high-frequency activity in MNs and understanding why this activity reaches MNs are relevant challenges to advance in the framework of MN-based CNS interfaces. Given the limitations of non-invasive neuroimaging technologies to record signals originating in the CNS, the interpretation of the high-frequency activity present in the MNs is complex, also because of the lack of reference signals guiding the analysis. By simultaneously recording brain and muscle activity, many previous studies have consistently proven that oscillatory activities at 20-30 Hz (beta rhythms) and 40-50 Hz (low-gamma rhythms) carried by MNs have a cortical origin[9,10,48–55]. This implies that, to some extent, muscles are driven by such cortical rhythms that MNs encode. For MNs to provide valuable information about the ongoing cortical activity that they receive, the characteristics of the corticospinal transmission of such signals must be stable across time. Also, transmission needs to be established with relatively small delays for MNs to be used to implement neural interfaces with the brain and modulations of cortical oscillatory inputs to MNs need to be strong enough to be clearly sensed from MNs.

Neurons in the corticospinal tract (the main descending pathway connecting the brain and the MNs) are entrained with ongoing activity in the neighbouring areas of the motor cortex, such as beta and gamma rhythms[56,57]. Even though highly heterogeneous and complex pathways connect cortical neurons and MNs, due to the distribution of transmission delays across the thousands of channels connecting the motor cortex to MNs, cortical neural activity is effectively transmitted to the muscles reliably through the fastest and most straightforward pathways[58]. This implies that MNs can provide a timely and stable readout of ongoing cortical neural activity. Fig. 5 shows an example of the linear association (measured using spectral coherence[59]) between cortical and MN frequency components (specifically beta rhythms, which are prominent during sustained contractions). The figure also shows that on average, there is a time delay of approximately 28 ms between cortical signals and MN activity, which corresponds to the expected transmission delays in the fastest pathways connecting the motor cortex and the tibialis anterior muscle (the measured muscle in this example). Overall, these results demonstrate the fast transmission of cortical rhythmic activity to the muscles, supporting the idea that MN decoding from muscle sensors enables interfacing with the motor cortex.

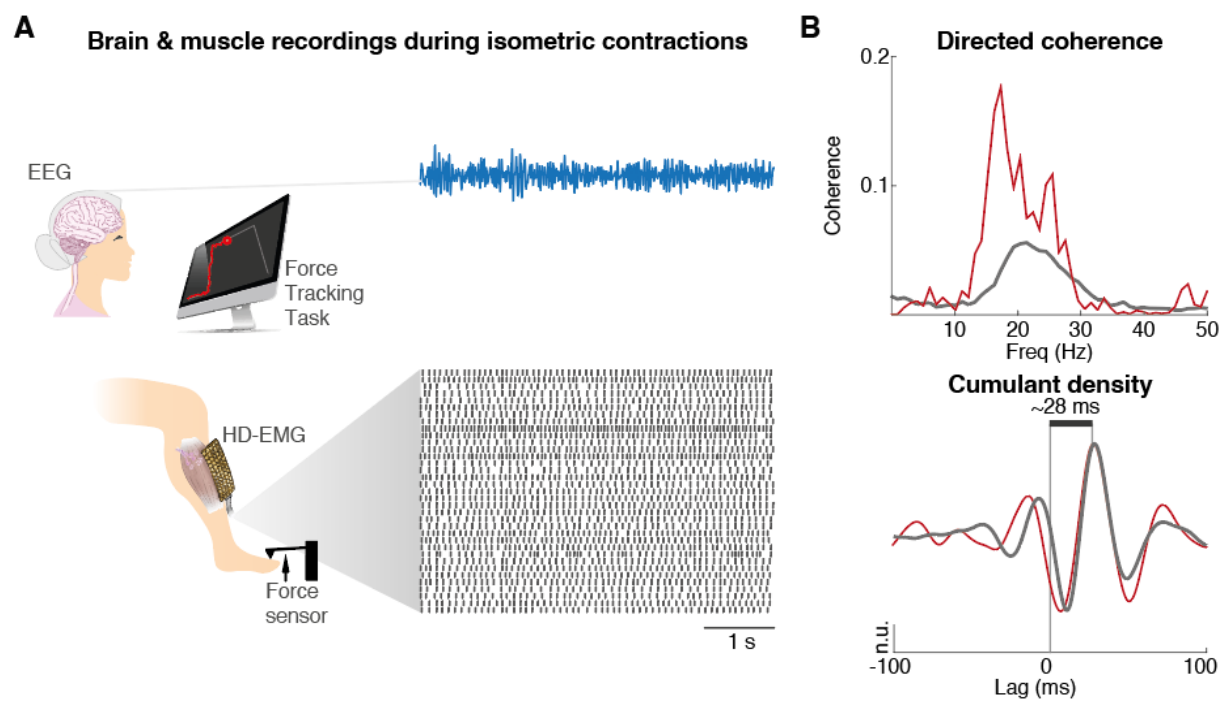

**Fig. 5. Stable transmission of cortical activity to muscles.** Simultaneous recordings of brain and muscle activity to assess their neural connectivity. (A) EEG from the primary motor cortex (blue trace) and HD-EMG from the tibialis anterior muscle were acquired while the ankle produced an isometric flexion force (approximately at 10% of the maximum force that the subject could generate). The MN activity decoded from the HD-EMG was analysed together with the EEG recordings from the scalp area above the cortical representation of the studied muscle. (B) Corticomuscular coherence in the descending (brain to muscle) direction (top panel) provides evidence for the linear transmission of cortical activity at ~20 Hz (vertical dashed line) to MNs. The analysis of the temporal association between the cortical and MN activity (bottom panel) indicates a lag between the correlated activity in the two recordings of ~28 ms, which corresponds to the average brain-muscle transmission delays through the fastest neural routes connecting the brain and the tibialis anterior[60]. The red traces indicate the data from one representative subject; the grey traces are the average group data (*n*= 19 subjects).

The use of muscle recordings to characterize the functional connection of the brain with MNs and as an indirect observation of brain signals can be further examined using neurofeedback paradigms. In a study by Carlowitz-Ghori et al., the authors used neuro-feedback paradigms to make subjects modify the coherence amplitudes in the beta band during sustained contractions without changing the elicited forces[61]. With regards to the use of MN activity to infer some ongoing brain signals, recent studies have used real-time decoding of EMG into MN activities to train subjects to volitionally modulate rhythmic beta activity (13-30 Hz) in the neural drive to muscles[62,63]. After a relatively short training process, this "muscle-based neuro-feedback" paradigm led to changes in beta activity detected in muscles and in cortical EEG. Results further indicated that beta events (or bursts) identified in the muscles during sustained contractions had similar characteristics (in terms of duration and rates of detection of bursts) to beta events identified from the motor cortex, suggesting that the beta activity transmitted by MNs has mainly a supraspinal origin. Continuing this line, investigating optimal ways to deliver feedback to subjects may allow us to gain a deeper insight into the characteristics of the high-frequency neural signals sampled by MNs, study how well subjects can learn to modulate high-frequency inputs to muscles during different types of motor tasks and generate new models explaining the functional roles of these types of inputs to muscles.

The studies described above are based on recordings of muscle activity during long periods of steady muscle contractions. Under such conditions, the observed fluctuations in cortical rhythmic activity projected to MNs are moderate, but strong enough to be detected in relatively small groups of MNs (for example, around 12 motor units were identified and tracked in real-time per subject in the neurofeedback experiments [31]). To further investigate how strongly the power of high-frequency (>10 Hz) cortical projections to MNs can fluctuate and how these changes can alter corticospinal transmission, it becomes relevant to use experimental paradigms that induce strong changes in high-frequency motor cortical oscillations during steady contractions[64–66]. A recent study assessed the magnitude and consistency of changes in the spectral contents of motor unit population activity during a movement preparation and cancellation paradigm[67]. Two types of trials were designed for this, 'GO' and 'NO-GO', guided by auditory cues. During the trials, subjects holding a tonic contraction of ankle dorsiflexion were first given a warning stimulus to prepare a motor response (ballistic contraction), and 1 s later they received an imperative stimulus that required them to either release the prepared action ('GO' trials) or to cancel the preparation, but keep holding the stable contraction ('NO-GO' trials). Cancellation of a prepared action is known to be associated with strong increases in cortical beta activity, which may have a role in preventing changes in motor cortical outputs[66,68]. The study showed that while participants successfully held the sustained contractions in the 'NO-GO' trials, salient changes in the power in different spectral bands in the MNs could be observed (see data from a representative subject in Fig. 6). Interestingly, part of these changes (between 13-30 Hz and 30-45 Hz) were related to changes in cortical activity taking place in parallel with the cancellation of the prepared action in the No-Go trials[67]. These results show evidence of successful corticospinal transmission of salient cortical oscillations in the context of not changing the motor outputs (which further supports the concept of MNs being able to simultaneously control force production and receive and be driven by high-frequency inputs from distal areas in the CNS). Further support of this idea can be found in a recent study providing consistent evidence of the corticomuscular transmission of beta bursts occurring without any detectable changes in the force produced by tonically contracted muscles (Fig. 6B) [69].

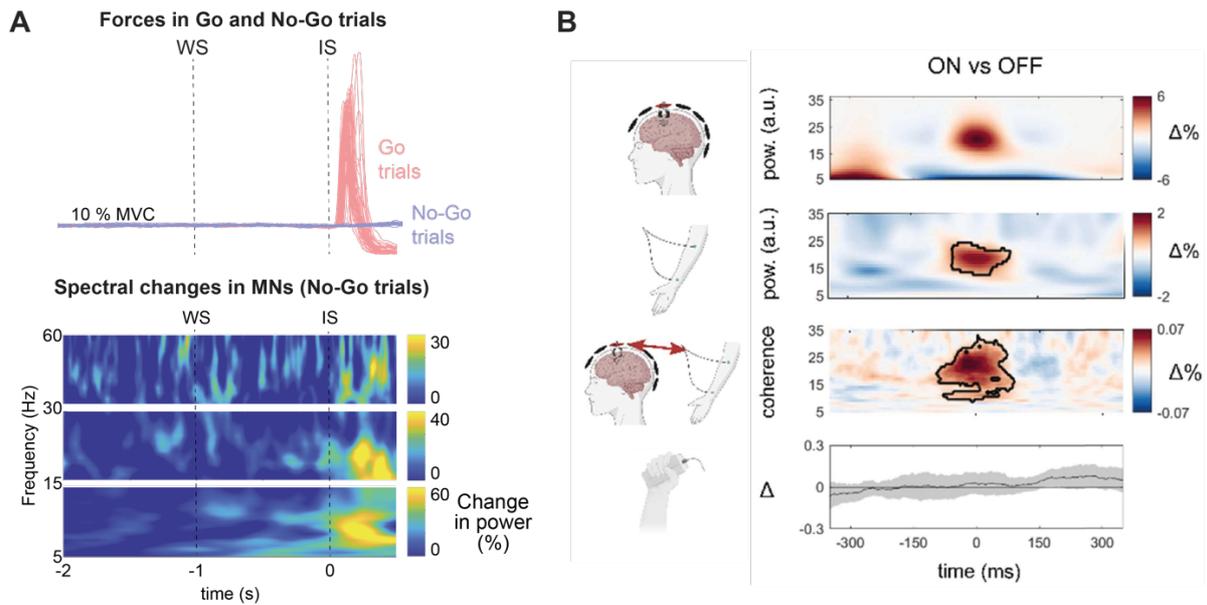

**Fig. 6. Evidence for changes in high-frequency contents sampled by MN that do not translate into force changes during isometric contractions.** (A) Changes in different frequency bands in the MN activity during the preparation and cancellation of a motor task in a Go/No-Go paradigm. (Top panel) Force traces from a representative subject in a study using a Go/No-Go paradigm to measure changes in high-frequency MN activity[67]. In each trial, subjects had to mildly activate the tibialis anterior muscle (producing a dorsiflexion force of 10% of the maximum force achievable by each subject). After some time keeping the force stable, subjects were presented two cues: a warning stimulus (WS) informed about the upcoming imperative stimulus (IS), which was given 1 s after the WS. The IS could either mean that subjects had to perform a fast dorsiflexion with the minimum reaction time possible (Go trials), or it could indicate to subjects that they had to remain in the same position without changing the produced force with the ankle (No-Go trials). (Bottom panel) Results from the same representative subject as the one used for the top panel. From this subject, 38 motor units could be decomposed and tracked across 46 valid No-Go trials. The composite spike train (summing the activity of all the motor units identified) was obtained and then a continuous wavelet transform analysis was used to analyse the time-frequency changes in the composite spike train signal. Results reflect important changes in the power at three frequencies (around 8, 20 and 46 Hz). In the average results across subjects, significant changes could only be found at frequencies above 15 Hz[67]. Colour scales were adapted to three frequency ranges (5-15 Hz, 15-30 Hz and 30-60 Hz). These results motivate the study of the high-frequency signals in MNs as a potential resource to indirectly study changes in the CNS. (B) Results from a study combining magnetoencephalography with EMG to analyse the cortico-muscular transmission of beta activity during sustained contractions[69]. The figure shows the averaged segments of brain recordings in which beta bursts are detected (top panel) and shows the corresponding activity in the muscles (second panel), which aligns with the beta modulations in the brain. The coherence measurements between brain and muscle signals (third panel) provide consistent evidence of the common origin of the beta activity measured in the brain and the muscle. Finally, the bottom panel shows how forces are kept unaltered during the periods of strong beta modulation at the brain and muscle levels (the plot shows the average and standard deviation of the measured forces across all trials considered to generate these plots).

The previous studies suggest that MN activity may allow us to distinguish brain states related to different types of mental processing. This was tested in a recent study in which MN activity was decoded during steady contractions while subjects performed different types of mental tasks (right foot motor imagery, imagining movements of both hands, mathematical mental operation, or not focusing on any specific tasks)[70]. Using a filter bank and a linear classifier, the mental tasks could be classified in some subjects with moderate to low accuracies, using only high-frequency components in the MN activity as input features for classification. Furthermore, the modulations in high-frequency components were incompatible with changes in force level. This study replicated classic brain-interfacing paradigms based on imagery with peripheral recordings, showing that some mental task-induced oscillations from supraspinal areas may leak down to MNs allowing the discrimination of mental tasks above chance level via EMG. Importantly, with respect to the other studies discussed above, this study also involved imagination of non-motor tasks, further supporting the notion that MN activity may be used to retrieve information about ongoing brain activity, even if it is not associated with motor processing.

Overall, there is consistent evidence that MNs receive and encode rhythmic activity that does not modulate forces and that have origins in the brain, *e.g.* in the motor cortex. To further advance in the interpretation and use of high-frequency neural inputs to MNs, it becomes critical that multiple common inputs to MNs can be separated. As previously discussed, this should be possible using standard spectral separation techniques if the different neural inputs to MNs are independent and have non-overlapping spectral bandwidths. However, this may not be expected if neural inputs have large bandwidths or if they are only transmitted to MNs through relatively small neural populations (in which case, the inputs will present harmonic components that can be summed with neural sources in other frequency bands, with non-linear mechanisms). An alternative possibility in this context may be to use other source separation methods that rely on the statistical properties of the inputs to be decoded and that may take advantage of the fact that inputs to MNs arising from different parts of the CNS may have different patterns of projections to MNs within a muscle or innervating groups of muscles[71,72]. Alternatively, spike prediction models may be implemented to estimate upstream neural activity from downstream spiking behaviour[73–75]. Interestingly, a recently proposed reinforcement learning-based point process framework has been used to generatively predict spike trains through behaviour-level rewards[76], i.e. using behavioural information to reduce the complexity of estimating neural activity from higher centres from downstream activity. The framework could be used to estimate neural information associated to movement from spiking activity of MNs coding cortical information and, as a consequence, the activity that does not influence movement.

*Prospects of the peripheral neural interfaces with the CNS*

The ability to decode CNS signals from MN activity presents an exciting opportunity to access neural information that would otherwise be difficult to obtain non-invasively in unconstrained contexts. This is especially important for understanding neural activity in deep brain structures and spinal circuits, which are not currently measurable with existing technologies but may project their activity onto MNs in specific motor contexts.

The possibility of using MNs as an indirect measure of the human brain is promising in the development of neural interfaces. Having access to large populations of MNs during a variety of motor conditions is expected to lead to improved estimates of high-frequency signals encoded by MNs and to a more refined understanding of how these signals co-variate with behaviour. Importantly, neural interfaces based on MN recordings may provide access to both spinal and brain signals (so long the connectivity between circuits in the CNS and the muscles is not disrupted by a lesion). With modern EMG recording and processing techniques, we can now track the activity of large volumes of MNs with increasing precision, generalizability across subjects and out of the lab, making this a viable approach for man-machine interfaces[20,22,77,78]. This type of interfacing can be accomplished with minimally- or even non-invasive devices, allowing for potentially large information transfer by measuring multiple neurons simultaneously and accurately. It also requires minimal calibration and may achieve long-term stability in tracking individual motor units across multiple sessions if the position of the EMG electrodes is kept similar relative to the measured muscles [79,80].

Using MNs to develop an interface with the human brain may be particularly interesting for expanding human motor abilities by allowing people to control artificial limbs concurrently with their natural limbs[81]. The difficulty in realizing such motor augmentation is to identify neural signals that can be controlled independently from and in coordination with natural motor control signals[82], and to transform the extracted information that is outside the motor space into a command that can allow a meaningful control of an electronic device[83]. MNs can be measured using wearable sensors and may transmit simultaneously low- and high-frequency neural contents to muscles, as discussed above. If parts of the high-frequency contents transmitted to muscles do not have any indirect links to motor function, they may represent a resource to expand the number of degrees of freedom that humans can control without interfering with the movements of natural limbs. This could enable us to extend the degrees of freedom that allow movement of the different parts in our body (see concept in Fig. 7).

As a step towards this goal, a recent study used real-time decomposition of MN activity to allow participants to control two parts of the spectral contents that MNs were encoding[62]. The study showed that after brief periods of training, subjects could partially dissociate the activity in the two frequency bands that they had to control independently. This provides a first proof of concept that it is possible to extend the degrees of freedom that a single muscle can control (from 1 degree to 2 in this case). To advance this line of research, work is now needed to identify methods to extract and classify high-frequency signals in MNs and develop optimal training protocols aimed at giving humans the ability to control neural processes originating in the brain and streaming through the muscles. One promising direction would consist in the data-driven extraction of latent variables from MN behaviour that are orthogonal to movement kinematics. This could be done using AI-based approaches applied to dynamic modelling and latent factor analysis in MN populations[84–86]. Approaches that enable the identification of behaviourally relevant subspaces from cortical activity could be adapted to dynamic systems at the spinal level using MN data decomposed from muscle recordings. This would allow for the separation of the behaviourally relevant and irrelevant latent factors that drive motor unit activity. Alternatively, neurofeedback on MNs may be used to train the CNS to increase the number of motor commands to muscles to reach augmented control[87].

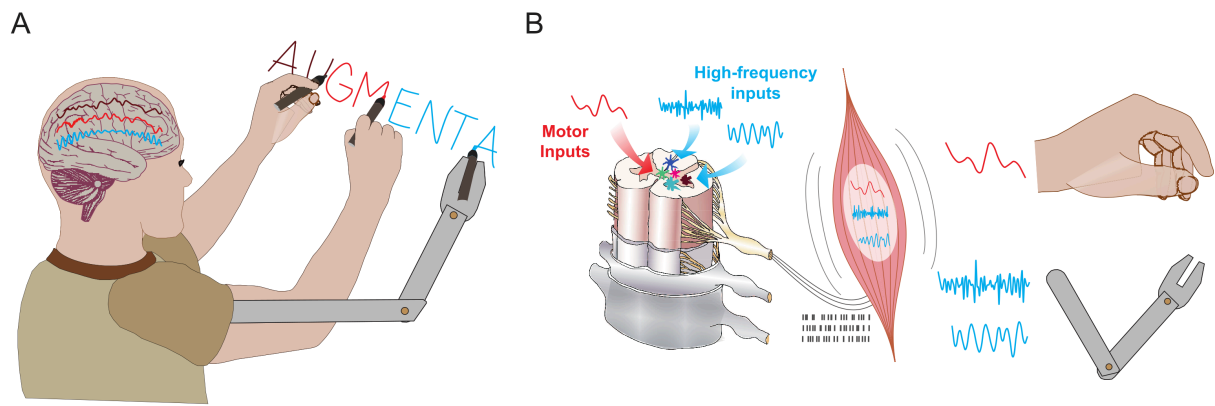

**Fig. 7. Human motor augmentation by extending the number of control signals extracted from human muscles.** (A) In the field of human motor augmentation, a key question relates to the extraction of neural information from the brain that allows humans to control artificial effectors. (Panel A is inspired by Stelarc´s third-hand art project; stelarc.org/?catID=20265). (B) The spinal cord projects multiple signals to the motor neuron (MN) pools innervating muscles. Inputs common to all MNs in a muscle and outside the motor band are linearly transmitted to muscles and are partly associated with ongoing brain activity. These high-frequency inputs to MNs and muscles may be used to augment the degrees of freedom that humans can use to control natural limbs and artificial effectors. In this view, the electric fields that can be measured over the human body reflect the activity of muscles which can be tracked back to cortical brain signals, thereby yielding a brain interface from peripheral sensors placed on limbs.

Finally, the possibility of decoding cortical activity from wearable sensors in the limbs to monitor brain states without a direct brain recording may also be used to refine existing interventions based on guiding neural changes in the CNS by stimulating it non-invasively[88,89]. MN-based information can be used in neuromodulation interventions, such as in closed-loop brain stimulation paradigms, where the information on the brain state that triggers the stimulation is obtained from a peripheral recording. This approach may encounter problems when the delay introduced by conduction to the periphery poses significant constraints on the closed-loop algorithm[90]. However, in applications where this delay is not a problem, such recordings would have several advantages with respect to current closed-loop brain stimulation approaches that estimate brain states directly from brain recordings[88]. First, electrical or magnetic stimulation of the CNS produces large artefacts that strongly affect brain recordings, but they have little impact on muscle recordings which should not affect the accuracy in decoding the spiking activity of MNs[91]. Second, oscillations at high frequencies can presumably be better identified from direct neural cell spike trains, as in the case of MN decoding, than from the interference EEG signals recorded from the scalp. Finally, if oscillations in a frequency band decoded from MNs are only caused by cortical projections, or if the contributions from different sources to a given frequency band of the spectral contents contained in the MN activity can be separated (*e.g.*, using separation methods described above in this perspective), then there would not be the need of separating multiple cortical sources simultaneously oscillating at overlapping frequencies. In fact, in this context and as discussed before, MNs are expected to provide reliable estimates of modulations in cortical rhythmic activity based on simulations of corticospinal projections to MNs.

*Considerations and limitations*

This perspective underscores recent strides in the analysis and interpretation of muscle recordings, aiming to motivate further exploration into harnessing the information encoded by MNs about CNS activity. The promise of muscle-based neural interfaces is evident in their practicality and potential for real-world applications, as demonstrated by recent studies[78]. However, some considerations are important regarding the current state of techniques and systems employed to extract MN information from muscles. Foremost among these considerations is the reliance of MN-based interfaces on muscle activation. Information at the entry points of MNs becomes accessible only upon their recruitment and firing. Furthermore, a critical challenge facing EMG-based MN recording methods is decoding MNs during dynamic contractions. As muscles and limbs initiate movement, the stability of motor unit action potential shapes diminishes, rendering linear decomposition methods unreliable for identifying and tracking single MNs. While recent literature suggests various approaches to address this limitation[18,22,24], non-invasive systems capable of tracking MNs during unrestricted movements are yet to be realized, constraining the applicability of MN-based neural interfaces beyond laboratory settings. Moreover, to accurately estimate neural inputs to MNs from their outputs during dynamic conditions, it is imperative to model neuromodulatory processes at the MN membrane reliably. Such processes may transiently disrupt the linear input-output curves of MNs. Although these fluctuations may not significantly impede the extraction of common inputs from a MN pool across different frequencies, they could introduce noise, compromising the fidelity with which modulations in various common inputs are tracked[37,92]. Finally, it is important to note that, at present, we still do not understand what are the functional roles of high-frequency inputs to MNs with cortical origin (*e.g.*, beta and gamma activity) and their link with motor control. Therefore, it is hard to predict the range of applications that may be developed by having access to this type of information from MNs. In order to address these challenging questions, further research on new ways to explore how MNs can give a handle on the high-frequency inputs is of great relevance.

**Conclusion**

In this perspective, we argue that electric fields generated by muscles in our bodies may provide an indirect observation of neural activity that originates in the central nervous system and is transmitted through motor neurons and muscles without directly modulating forces. They reflect the behaviour of neurons that constitute the output layer of the neural network formed by the nervous system. Since at least part of the information that arrives at the output of the nervous system is known to originate in regions of the central nervous system like the cortex, recording from muscle tissue may allow us to establish a novel type of human interfacing not only with the peripheral nervous system, as discussed in previous works, but also with the CNS, as argued in this review. The presence of cortical oscillations in the activity of MNs opens new frontiers of interfacing with applications that range from neurofeedback to human movement augmentation.

**Funding**

This work was partly funded by the EC H2020 grants NIMA (FETOPEN 899626) and Natural BionicS (ERC Synergy 810346), UKRI's EPSRC grant EP/T020970/1 and BBSRC grant BB/V00896X/1. JI is supported by a Ramón y Cajal grant (RYC2021-031905-I) funded by MCIN/AEI/10.13039/501100011033 and UE's NextGenerationEU/PRTR funds, by the European Research Council (ERC) through the Horizon Europe 2022 Starting Grant program under REA grant agreement number 101077693 and by MICIU/AEI and FEDER, UE (Grant PID2022-138585OA-C32).

**Author contributions**

JI and DF drafted the manuscript. All authors contributed to literature review, conceptualization and editing of the manuscript.

**Competing interests**

The authors declare no competing interests.

**References (max. 100):**